\documentclass[12pt,preprint]{aastex}

\shorttitle{Tests for Standard Disk Models of AGNs}

\shortauthors{Liu et al.}

\begin{document}

\title{Tests for Standard Accretion Disk Models by Variability in Active Galactic Nuclei}

\author{H. T. Liu\altaffilmark{1}, J. M. Bai\altaffilmark{1}, X. H. Zhao\altaffilmark{1}, and L. Ma\altaffilmark{2}}

\altaffiltext{1}{National Astronomical Observatories/Yunnan
Observatory, the Chinese Academy of Sciences, Kunming, Yunnan
650011, China; liuhongtao1111@hotmail.com; baijinming@ynao.ac.cn.}

\altaffiltext{2}{Physics Department, Yunnan Normal University,
Kunming 650092, China.\\
send offprint requests to H. T. Liu}

\begin{abstract}

In this paper, standard accretion disk models of AGNs are tested
using light curves of 26 objects well observed for reverberation
mapping. Time scales of variations are estimated by the most
common definition of the variability time scale and the
zero-crossing time of the autocorrelation function of the optical
light curves for each source. The measured time scales of
variations by the two methods are consistent with each other. If
the typical value of the viscosity parameter $\alpha \sim 0.1$ is
adopted, the measured optical variability time scales are most
close to the thermal time scales of the standard disks. If
$\alpha$ is allowed to range from $\sim 0.03$ to $\sim 0.2$, the
measured time scales are consistent with the thermal time scales
of the standard disks. There is a linear relation between the
measured variability time scales and black hole masses, and this
linear relation is qualitatively consistent with expectation of
the standard accretion disk models. The time lags measured by the
ZDCF between different bands are on the order of days. The
measured time lags of NGC 4151 and NGC 7469 are marginally
consistent with the time lags estimated in the case of continuum
thermal reprocessing for the standard accretion disk models.
However, the measured time lags of NGC 5548 and Fairall 9 are
unlikely to be the case of continuum thermal reprocessing. Our
results are unlikely to be inconsistent with or are likely to be
conditionally in favor of the standard accretion disk models of
AGNs.

\end{abstract}


\section{INTRODUCTION}
Large flux variations on time scales from years to hours are
common in active galactic nuclei (AGNs), and longer time scale
variations of the order of months to years may be related to the
propagation of the shorter time scale variations (e.g., Ulrich et
al. 1997). The combination of high flux variability and short
variability time scales implies that the energy conversion in AGNs
is more efficient than the ordinary stellar processes. Accretion
of matter onto a black hole can have high energy release
efficiency (Rees et al. 1982; Rees 1984). The evidence that AGNs
such as quasars and Seyfert galaxies are powered by gravitational
accretion of matter onto supermassive black holes is now quite
convincing. Certainly there has been no definitive detection of
the relativistic effects that would be required for unambiguous
identification of a singularity, although studies of the iron
K$\alpha$ emission line in the X-ray spectra of AGNs currently
provides some promise (e.g., Reynolds \& Nowak 2003).

In general optical--UV radiations of most non-blazar type sources
are within the so-called Big Blue Bump. The optical variability is
characterized as poorly understood, but is nevertheless recognized
as a means of probing physical scales that cannot be resolved
spatially by any telescope or instrument (e.g., Netzer \& Peterson
1997; Peterson et al. 2004; Wold et al. 2007). A number of models
have been proposed to explain optical--UV quasar variability. One
way of attempting to help constraining the proposed models is to
find relationships between variability and other parameters of
AGNs, such as black hole mass. The black hole mass is a
fundamental parameter of AGNs, and the discovery of such a
relationship -- or lack thereof -- may provide useful clues to the
physical mechanisms behind the variability (e.g., Wold et al.
2007). Processes intrinsic to the central engine itself could
dominate. Wold et al. (2007) investigate the dependence of quasar
variability on black hole mass, and find that black hole mass
correlates with the measured variability amplitude. A number of
models for quasar optical variability exist but there are no clear
predictions relating black hole mass and variability amplitude.
Different sources of optical variations can be associated with
different characteristic time scales, and many of these time
scales depend on black hole mass. Collier \& Peterson (2001)
attempt to define a relationship between black hole mass and
characteristic variability time scale. Studying 10 well-monitored
AGNs, they report evidence of black hole mass correlating with
characteristic optical variability time scales that are roughly
consistent with accretion disk thermal time scales.

The standard accretion disk is the basic model for a radiatively
efficient, geometrically thin, optically thick disk (Shakura \&
Sunyaev 1973). In the standard picture this accretion disk
radiates thermally mainly in the optical/UV bands for AGNs with
black hole masses of $\sim 10^6$--$10^9 \/\ M_{\sun}$. AGNs with
black hole masses of $\sim 10^7$--$10^9 \/\ M_{\sun}$ would be
expected to have accretion disk thermal characteristic time scales
of the order of months to years. Many investigations based on
central radiations from thin accretion disks have been done (e.g.,
Ebisawa et al. 1991; Hanawa 1989; Li et al. 2005; Pereyra et al.
2006; Zimmerman et al. 2005). Connections of jets and disks, a
very important aspect of AGN researches, have been investigated on
the basis of standard accretion disks (e.g., Meier 2001, 2002).
Though many investigations are on the basis of standard disks,
only a few investigations aim at testing standard accretion disk
models by observations. Collier et al. (1998) briefly discussed
the relation of time delays between the UV and optical continuum
variations with accretion disk in NGC 7469. It is believed for
non-blazar type AGNs that the optical/UV emissions are produced
thermally from accretion disks. The radiation energies of thermal
emissions emitted in accretion disks are from two possible
contributions. One well-known origin is the local viscous
dissipation in accretion disks. This local viscous dissipation can
produce the local thermal equilibrium, and then the local
blackbody emissions (e.g., Krolik 1999). Another origin is the
reprocessed X-rays. The X-rays are commonly attributed to Compton
up-scattering of the thermal UV photons produced by the viscous
dissipation (e.g., Sunyaev \& Titarchuck 1980; Haardt \& Maraschi
1991). In the case of thermal emissions from viscous dissipation,
the accretion flow fluctuations travelling inwards across the
emitting regions affect first the optical emitting region at outer
radii, and then the UV emitting region at inner radii. Then the
longer wavelength variations are likely to lead the shorter
wavelength ones. If the radial temperature profiles of accretion
disks are not set primarily by viscous effects, but by irradiation
from the central X-ray sources, the longer wavelength variations
are likely to lag the shorter wavelength ones for thermal
emissions from continuum thermal reprocessing. The flux
variability must occur on a physical time scale that is consistent
with the chosen model. The time scales of interest are the light
crossing, dynamical, thermal, and sound crossing time scales that
are set by the black hole mass (Frank, King, \& Raine 2002), and
the order-of-magnitude scales are
\begin{equation}
\tau_{\rm{l}}=6M_{8}\xi_{3}\/\ \rm{days},
\end{equation}
\begin{equation}
\tau_{\rm{dyn}}=6M_{8}\xi_{3}^{3/2}\/\ \rm{months},
\end{equation}
\begin{equation}
\tau_{\rm{th}}=\tau_{\rm{dyn}}/ \alpha=5M_{8}\xi_{3}^{3/2}\/\
\rm{yrs},
\end{equation}
\begin{equation}
\tau_{\rm{s}}=70M_{8}\xi_{3}T^{-1/2}_{5} \/\ \rm{yrs},
\end{equation}
where $M_8=M_{\rm{BH}}/10^8 \/\ M_{\sun}$, $\alpha$ ($\sim 0.1$)
is the Shakura-Sunyaev viscosity parameter (Shakura \& Sunyaev
1973), $T_5=T/10^5 \/\ \rm{K}$, and $\xi_3=r_{\rm{d}}/10^3 \/\
r_{\rm{g}}$ ($r_{\rm{d}}$ is the disk radius, and $r_{\rm{g}}=G
M_{\rm{BH}}/c^2$ the gravitational radius). In order to determine
which physical mechanism is responsible for the variability and to
test standard accretion disk models, it is necessary to connect
the observed variability time scales with one of the above
physical time scales and to search the correlation of black hole
masses with characteristic optical variability time scales, and it
is important to compare the observed time lags between different
bands with the theoretical values predicted by the standard
accretion disk models.

The structure of this paper is as follows. The sample and data are
in $\S$ 2. The calculations of temperature profiles are described
in $\S$ 3. $\S$ 4 presents variability time scales and time lags.
$\S$ 4.1 is analysis of variability time scale, $\S$ 4.2 analysis
of time lag, and $\S$ 4.3 comparison to models. The last section
is discussions and conclusions. Throughout this paper, we use a
flat cosmology with a deceleration factor $q_0=0.5$ and a Hubble
constant $H_0=75 \/\ \rm{km \/\ s^{-1} \/\ Mpc^{-1}}$.

\section{THE SAMPLE AND DATA}

The objects listed in Table 1 are based on the samples analyzed by
Kaspi et al. (2000) and Peterson et al. (2004), but the light
curve data comes from a variety of sources. The rest frame
wavelengths and references of light curves are listed in columns
(3) and (4) of Table 1, respectively. The optical variability time
scales are estimated by the light curves around 5100 $\rm{\AA}$.
There are four objects, Fairall 9, NGC 4151, NGC 5548, and NGC
7469, that have multi-wavelength light curves well observed at the
optical/UV bands. The multi-wavelength light curves are used to
estimate time lags for the four objects.

The black hole masses for AGNs have been well estimated by the
reverberation mapping technique (e.g., Kaspi et al. 2000, 2005;
Peterson et al. 2004, 2005; Vestergaard \& Peterson 2006). The
masses of the central black holes of quasars span a large range of
$10^7 \/\ M_{\sun} \lesssim M_{\rm{BH}}\lesssim 3\times 10^9 \/\
M_{\sun}$, and have an upper limit of $M_{\rm{BH}}<10^{10} \/\
M_{\sun}$ (McLure \& Dunlop 2004; Vestergaard 2004). The black
hole masses used in this paper are taken from Peterson et al.
(2004), and are listed in column (5) of Table 1. The bolometric
luminosity $L_{\rm{bol}}$ of objects except for Mrk 279 are take
from Woo \& Urry (2002), and are listed in column (6) of Table 1.
The bolometric luminosity of Mrk 279 is estimated by
$L_{\rm{bol}}\approx 9\lambda L_{\lambda}(5100\/\ \rm{\AA})$
(Kaspi et al. 2000), with $\lambda L_{\lambda}(5100\/\ \rm{\AA})$
taken from Peterson et al. (2004).

\section{CALCULATIONS OF TEMPERATURE PROFILES}

The local effective temperatures of accretion disks are functions
of radii $r_{\rm{d}}$, black hole mass $M_{\rm{BH}}$, spin
$a_{\ast}$, and mass accretion rate $\dot{M}$ (e.g., Ebisawa et
al. 1991; Hanawa 1989; Kubota et al. 2005; Li et al. 2005; Pereyra
et al. 2006; Shakura \& Sunyaev 1973; Zimmerman et al. 2005). The
standard accretion disk is the basic model for a radiatively
efficient, geometrically thin disk. If the central black holes are
Kerr ones, the local effective temperature of the standard disk is
given in the Kerr metric as (Krolik 1999)
\begin{equation}
T_{\rm{eff}}(X_{\rm{d}})=\left[ \frac{3G M_{\rm{BH}} \dot{M}}{8
\pi \sigma_{\rm{SB}} r_{\rm{g}}^3 X_{\rm{d}}^3}
R_{\rm{R}}(X_{\rm{d}})\right] ^{1/4},
\end{equation}
where $\sigma_{\rm{SB}}$ is the Stefan-Boltzmann constant, $G$ is
the gravitational constant, $\dot{M}$ is the mass accretion rate
of the central black hole, $M_{\rm{BH}}$ is the central black hole
mass, and $X_{\rm{d}}=r_{\rm{d}}/r_{\rm{g}}$ is the disk radius in
units of the gravitational radius $r_{\rm{g}}$. The function
$R_{\rm{R}}(X_{\rm{d}})$ in equation (5) is defined as
\begin{equation}
R_{\rm{R}}(X_{\rm{d}})=\frac{C(X_{\rm{d}})}{B(X_{\rm{d}})},
\end{equation}
where the functions $B(X_{\rm{d}})$ and $C(X_{\rm{d}})$ are,
respectively, (Krolik 1999)
\begin{equation}
B(X_{\rm{d}})=1-\frac{3}{X_{\rm{d}}}+\frac{2a_{\rm{\ast}}}{X_{\rm{d}}^{3/2}},
\end{equation}
and
\begin{eqnarray}
C(X_{\rm{d}}) &=&
1-\frac{y_{\rm{ms}}}{y}-\frac{3a_{\rm{\ast}}}{2y}\ln\left(\frac{y}{y_{\rm{ms}}}\right)
 \nonumber\\
 &&-\frac{3(y_1-a_{\rm{\ast}})^2}{yy_1(y_1-y_2)(y_1-y_3)}\ln\left(\frac{y-y_1}{y_{\rm{ms}}-y_1}\right)
 \nonumber\\
 && -\frac{3(y_2-a_{\rm{\ast}})^2}{yy_2(y_2-y_1)(y_2-y_3)}\ln\left(\frac{y-y_2}{y_{\rm{ms}}-y_2}\right)
 \nonumber\\
 && -\frac{3(y_3-a_{\rm{\ast}})^2}{yy_3(y_3-y_1)(y_3-y_2)}\ln\left(\frac{y-y_3}{y_{\rm{ms}}-y_3}\right) \;,
\end{eqnarray}
where $y=\sqrt{X_{\rm{d}}}$, $a_{\rm{\ast}}=cJ/GM^2_{\rm{BH}}$ is
the dimensionless spin parameter of the central black hole with
the spin angular momentum $J$, $y_{\rm{ms}}=\sqrt{X_{\rm{ms}}}$ is
the value of $y$ at the marginally stable orbit, and $y_{1,2,3}$
are the three roots of $y^3-3y+2a_{\rm{\ast}}=0$, respectively
(e.g., Krolik 1999; Reynoldsa \& Nowakb 2003).

Assuming prograde orbits, the radii of the marginally stable
orbits in the equatorial plane of a Kerr black hole are (Bardeen
et al. 1972)
\begin{equation}
X_{\rm{ms}}=3+Z_2-\left[(3-Z_1)(3+Z_1+2Z_2) \right]^{1/2},
\end{equation}
where
\begin{equation}
Z_1=1+\left(1-a^2_{\rm{\ast}}\right)^{1/3}\left[\left(1+a_{\rm{\ast}}\right)^{1/3}+
\left(1-a_{\rm{\ast}}\right)^{1/3}\right],
\end{equation}
and
\begin{equation}
Z_2=\left(3a^2_{\rm{\ast}}+Z_1^2\right)^{1/2}.
\end{equation}
The marginally stable orbits in the equatorial plane correspond to
the maximum efficiency of energy release as a result of accretion,
assuming prograde orbits (Kembhavi \& Narlika 1999)
\begin{equation}
\eta_{\rm{max}}=1-\frac{X_{\rm{ms}}-2+a_{\rm{\ast}}X_{\rm{ms}}^{-1/2}}{\sqrt{X_{\rm{ms}}\left(
X_{\rm{ms}}-3+2a_{\rm{\ast}}X_{\rm{ms}}^{-1/2}\right)}}.
\end{equation}
According to the defination of the efficiency $\eta$ with which
various types of black holes convert rest mass-energy into
outgoing radiation (Thorne 1974), the mass accretion rate of the
central black hole can be estimated by the formula
\begin{equation}
\dot{M}=\frac{L_{\rm{bol}}}{\eta_{\rm{max}}c^2}.
\end{equation}

The dimensionless spin parameter of a black hole can take on any
value in the range $-1\leq a_{\rm{\ast}} \leq 1$, where negative
values of $a_{\rm{\ast}}$ correspond to a black hole that
retrogrades relative to its accretion disk. For simplicity we
consider only prograde spins up to the Thorne spin equilibrium
limit, i.e. $0\leq a_{\rm{\ast}}\leq 0.998$ (Thorne 1974). The
limiting value of $a_{\rm{\ast}}=0.998$ for black hole spins was
first discussed in Thorne (1974). Recent work on
magnetohydrodynamic accretion disks suggest a rather lower
equilibrium spin (e.g., Gammie et al. 2004; Krolik et al. 2005).
It is suggested that spin equilibrium is reached at
$a_{\rm{\ast}}\approx 0.93$ through accretion of gases onto the
central black holes, and mergers of black holes with comparable
mass can result in a final spin of $a_{\rm{\ast}}\sim $ 0.8--0.9
(Gammie et al. 2004). Krolik et al. (2005) suggested that
equilibrium spins as low as $a_{\rm{\ast}}\sim 0.9$ are within the
realm of possibility. Brenneman \& Reynolds (2006) obtained a
formal constraint on the dimensionless black hole spin parameter
of $a_{\rm{\ast}}=0.989^{+0.009}_{-0.002}$ at 90\% confidence for
the Seyfert galaxy MCG--06-30-15. A value of
$a_{\rm{\ast}}=0.9939^{+0.0026}_{-0.0074}$ for the Galactic Center
black hole is obtained by Aschenbach et al. (2004). Considering
the probable ranges of spin parameter $a_{\rm{\ast}}$ suggested
above, we take four values of spin parameter $a_{\rm{\ast}}=0.5$,
$0.8$, $0.9$, and $0.998$ in the Kerr metric to calculate the
temperature profiles. Combining equations (5)--(13) and the
parameters of $M_{\rm{BH}}$, $L_{\rm{bol}}$, and $a_{\rm{\ast}}$,
we can calculate the surface effective temperature profiles.

The local effective temperature in equation (5) is arrived by
assuming local thermal equilibrium (LTE) in disk. A consequence of
the LTE radiation assumption is that specific photon frequency
$\nu$ maps to specific radius $r_{\rm{\nu}}$ in the disk.
According to discussion of Krolik (1999, see equation (7.53) in
page 155), most of the light at frequency $\nu$ is emitted near
the radius $r_{\rm{\nu}}$ for the local blackbody. Then the
optical spectrum may still be dominated by emission from the outer
radii, and the UV spectrum can be dominated by emission from the
inner radii. So variations in the outer disk might manifest
themselves significantly in the observed spectrum.

\section{VARIABILITY TIME SCALE AND TIME LAG}
Two methods are applied to analysis of variability time scale. One
is the most common definition of the variability time scale (e.g.,
Wagner \& Witzel 1995). Another is a well defined quantity, the
zero-crossing time of the autocorrelation function of light
curves. Time lags are analyzed by the z-transformed discrete
correlation function (ZDCF; Alexander 1997). Then the analysis
results are compared to predications of accretion disk models.
\subsection{Analysis of Variability Time Scale}
The variability time scales have been defined in different ways.
The most common definition of the variability time scale
$\tau=F/|\Delta F/\Delta t|$ and the more conservative approach of
$\tau=|\Delta t/\Delta \ln F|$ have the advantage of weighting
fluctuations by their amplitudes, where $F$ is the flux, and
$\Delta F$ is the variability amplitude in the time scale $\Delta
t$ (e.g., Wagner \& Witzel 1995). Here we use the most common
definition of variability time scale $\tau=F/|\Delta F/\Delta t|$,
where $F$ is taken as the flux at the minimum. In this paper, we
refer to the interval between subsequent local minima and maxima
at the adjacent valleys and peaks in the entire light curve.
First, we select subsequent valley and peak sufficiently dense
sampled in one light curve. Second, variations of $\Delta F/F \ge
30\%$ between the subsequent minimum and maximum are required
within the time scale $\Delta t$. The estimated values of $\tau$
are listed in column (2) of Table 2. The uncertainty on the values
of $\tau$ are estimated by the relation $\sigma_{\tau}=\Delta
t(\sigma_{F_{\rm{min}}}|\Delta
F|-F_{\rm{min}}|\sigma_{F_{\rm{max}}}-\sigma_{F_{\rm{min}}}|)/|\Delta
F|^2$, where $\Delta F=F_{\rm{max}}-F_{\rm{min}}$,
$\sigma_{F_{\rm{max}}}$ is the observed error of $F_{\rm{max}}$,
and $\sigma_{F_{\rm{min}}}$ is the observed error of
$F_{\rm{min}}$.

For most AGNs, it is difficult to define a single characteristic
variability time scale. One approach to a single time scale is
described by Giveon et al. (1999). Their definition is given as
the zero-crossing time of the autocorrelation function (ACF). If
there is an underlying signal with a typical variability time
scale in the light curve, the width of the ACF peak near zero time
lag will be proportional to this variability time scale (e.g.,
Giveon et al. 1999; Netzer et al. 1996). This zero-crossing time
of the ACF, $\tau_{\rm{0}}$, is a well defined quantity, and is
used as a characteristic variability time scale (e.g., Alexander
1997; Giveon et al. 1999; Netzer et al. 1996). Another function
used in variability studies to estimate the variability time scale
is the first-order structure function (SF) (e.g., Trevese et al.
1994). There is a simple relation between the ACF and the SF (see
Eq. (8) in Giveon et al. 1999). Therefore only an ACF analysis is
performed on our light curves. Comparison of $\tau$ with $\tau_0$
is performed to test the reliability of the variability time scale
$\tau$ listed in column (2) of Table 2. The ACF is estimated by
the ZDCF (Alexander 1997). It has been shown that this method is
statistically robust even when applied to very sparsely and
irregularly sampled light curves (Alexander 1997). The ZDCF was
calculated for all of the light curves used to estimate $\tau$.
Following Giveon et al. (1999), a least-squares procedure is used
to fit a fifth-order polynomial to the ZDCF, and the ZDCF fit is
used to evaluate the zero-crossing time in the observer's frame.

The evaluated results are listed in column (3) of Table 2. For one
light curve, the ZDCF code of Alexander (1997) can automatically
set how many bins are given and used to calculate the ACF. Thus,
the time lag and its uncertainty are immediately given for each
bin in the ACF. However, this code cannot estimate the uncertainty
on the fit value of $\tau_0$ to the ACF. If the fit $\tau_0$ is
most near the time lag of one bin in the ACF, the uncertainty of
the fit $\tau_0$ may be approximated by the uncertainty of time
lag in this bin in the ACF. Thus, the uncertainty on the values of
$\tau_0$ in Table 2 is assumed to be the errors of the ACF points
nearest to the fit values of $\tau_0$.

For comparison, we plotted $\tau$ versus $\tau_0$ in Figure 1. It
can be seen in Figure 1 that the data points are basically shared
by two sides of the line $\tau_0=\tau$. The linear regression
analysis shows that there is a correlation between $\tau$ and
$\tau_0$ with Pearson correlation coefficient $r=0.766$ at the
chance probability $P=5.1\times 10^{-6}$. The regression line
fitted by the ordinary least-squares bisector regression analysis
(Isobe et al. 1990) is
\begin{equation}
\tau_{\rm{0}}/(1+z)=-96.1(\pm33.8)+1.5(\pm0.3)\tau/(1+z),
\end{equation}
where $z$ is the redshift, and $\tau$ and $\tau_0$ are in units of
$\rm{days}$. This suggests that the $\tau$ and $\tau_0$ are
acceptable to characterize the typical variability time scale, and
that the estimated results of $\tau$ listed in column (2) of Table
2 are reliable.

\subsection{Analysis of Time Lag}
Cross-correlation function (CCF) analysis is a standard technique
in time series analysis to find time lags between light curves at
different wavelengths, and the definition of the CCF assumes that
the light curves are uniformly sampled. However, in most cases the
sampling is not uniform. The interpolated cross-correlation
function (ICCF) method of Gaskell \& Peterson (1987) uses a linear
interpolation scheme to determine the missing data in the light
curves. On the other hand, the discrete correlation function (DCF;
Edelson \& Krolik 1988) can utilize a binning scheme to
approximate the missing data. Apart from the ICCF and DCF, there
is another method of estimating the CCF in the case of
non-uniformly sampled light curves, that is, the z-transformed
discrete correlation function (Alexander 1997). The ZDCF was used
as an estimation of the ACF in $\S$4.1; here it is used as an
estimation of the CCF. The ZDCF is a binning type of method as an
improvement of the DCF technique, but has a notable feature that
the data are binned by equal population rather than equal bin
width $\Delta \tau$ as in the DCF. It has been shown in practice
that the calculation of the ZDCF is more robust than the ICCF and
the DCF when applied to sparsely and unequally sampled light
curves (e.g., Edelson et al. 1996; Giveon 1999; Roy et al. 2000).
The ZDCF is calculated in this paper.

In general, it seems to be true that the time lag is better
characterized by the centroid $\tau_{\rm{cent}}$ of the DCF and
the ICCF rather than by the peak $\tau_{\rm{peak}}$, namely, the
time lag where the linear correlation coefficient has its maximum
value $r_{\rm{max}}$ (e.g., Peterson et al. 2004, 2005).
$\tau_{\rm{peak}}$ is much less stable than $\tau_{\rm{cent}}$ in
both the DCF and the ICCF, and $\tau_{\rm{peak}}$ is much less
stable in the DCF than in the ICCF (Peterson et al. 2005). Then we
prefer that the time lag estimated from the ZDCF method is
characterized by the centroid $\tau_{\rm{cent}}$ of the ZDCF, for
the ZDCF is an improvement of the DCF method. The centroid time
lags $\tau_{\rm{cent}}$ are computed using all points with
correlation coefficients $r\ge 0.8 r_{\rm{max}}$, and the
uncertainties for time lags of data points in the ZDCF are
computed with a large number ($1,000$) of Monte Carlo
realizations. The ZDCFs of four objects are presented in Figures
2--5, and the measured time lags are listed in column (4) of Table
3.

\subsection{Comparison to Models}
There is a correlation between the black hole mass $M_8$ and the
measured characteristic variability time scale $\tau$ with Pearson
correlation coefficient $r=0.760$ at the chance probability
$P=6.6\times 10^{-6}$ (see Fig. 6). The regression lines fitted by
the bisector regression analysis are
\begin{equation}
\tau/(1+z) = 0.27(\pm0.04)+0.12(\pm0.02)M_{8} \/\ \rm{yrs}.
\end{equation}If $r_{\rm{d}}\sim 100 \/\ r_{\rm{g}}$ in equation (3)
with viscosity parameter $\alpha=0.1$, there is a relation of
$\tau_{\rm{th}}\sim 0.15 M_8$. Though the intercept in equation
(15) differs from the intercept predicted by equation (3), this
predicted slope of $\sim 0.15$ is consistent with the one in
equation (15). This indicates that the linear correlation between
black hole mass and characteristic variability time sale is
qualitatively consistent with the expectation of equation (3) that
the thermal time scale is essentially linearly related with the
black hole mass. Thus, equation (15) is qualitatively consistent
with expectations of the standard accretion disk models.

According to the standard accretion disk models, the optical/UV
emissions are produced thermally in accretion disks. The standard
accretion disk models are used to estimate the radii of maximum
optical/UV emissions. We consider accretion disk to be composed of
rings with approximately uniform temperature radiating locally as
blackbody, and estimate the radii of maximum flux emission at
different wavelengths using a disk radial temperature profile
given by equation (5). Then the light crossing, dynamical,
thermal, and sound crossing time scales are estimated by equations
(1)--(4), respectively, assuming viscosity parameter $\alpha =
0.1$. The calculated results are presented in columns (4)--(7) of
Table 2, respectively. It can be seen from columns (2)--(7) of
Table 2 that the thermal time scales are most close to the optical
variability time scales among the four physical time scales, but
the light crossing and dynamical time scales are much smaller than
the measured time scales of optical variations, and the sound
crossing time scales are much larger than the measured time
scales. This might indicate that the optical variations result
from the thermal instability in accretion disks or one mechanism
related to it. Though, it cannot be affirmed that the optical
variations result from the accretion disk thermal instability, the
linear relation presented in equation (15) is qualitatively
consistent with expectation of equation (3) that the thermal time
scale is essentially linearly related with the black hole mass.
These above results are obtained by adopting the viscosity
parameter $\alpha = 0.1$ for each source in our sample. In
practice, various values of $\alpha$ are suggested and used in
investigations (e.g., Afshordi \& Paczynski 2003; Khajenabi \&
Shadmehri 2007; Merloni 2003; Merloni \& Nayakshin 2006; Pariev et
al. 2003). If the viscosity parameter $\alpha$ is allowed to range
from $\alpha \sim 0.03$ to $\alpha \sim 0.2$ (e.g., Afshordi \&
Paczynski 2003), calculations show that the combinations of
$\alpha \sim 0.03$--$0.2$ and $a_{\ast}=0.5$--$0.998$ can result
in the thermal time scales that are in well agreement with the
optical variability time scales presented in Table 2. Then it is
likely that the optical variations result from the accretion disk
thermal instability.

The radiation energies emitted in accretion disks are probably
from the continuum thermal reprocessing and/or the local viscosity
dissipation (e.g., Ulrich et al. 1997). If the X-rays illuminating
optically thick material in thin disk produce the optical--UV
emissions through thermal reprocessing, the optical and UV
variations following the X-ray variations are probably correlated
with the UV variations leading the optical ones. The time lags in
the case of continuum thermal reprocessing are estimated for the
standard accretion disks with black hole spin parameter
$a_{\ast}=0.5$, $0.8$, $0.9$, and $0.998$. The relevant time lags
are listed in column (5) of Table 3. The plus signs of values in
column (5) mean that the variations at longer wavelengths lag the
variations at shorter wavelengths. It can be seen from columns (4)
and (5) of Table 3 that the measured time lags are marginally
consistent with those predicted by the standard accretion disks
for NGC 4151 and NGC 7469. This implies that the optical and UV
emissions are likely to be the reprocessed X-rays for NGC 4151 and
NGC 7469. In addition, the time lags decrease slightly as spin
parameter $a_{\ast}$ increases. For Fairall 9 and NGC 5548, the
signs of the continuum thermal reprocessing time lags are contrary
to those of the measured time lags. This indicates that the
optical/UV emissions are unlikely to be the reprocessed X-rays for
Fairall 9 and NGC 5548. If variations in the accretion flow affect
first the flux at outer radii, and then in the inner region, this
maybe result in correlated optical/UV light curves with longer
wavelength variations leading shorter wavelength variations. As a
reference time scale, the sound crossing time in a standard
accretion disk between these radii are estimated for Fairall 9 and
NGC 5548 by adopting $a_{\ast}=0.998$. The estimated results are
listed in column (6) of Table 3. The minus signs of values in
column (6) mean that the variations at outer radii lead the
variations at inner radii. It can be seen from columns (4) and (6)
of Table 3 that the measured time lags are much smaller than those
predicted by the standard accretion disks in the case of accretion
flow fluctuations travelling inwards.

\section{DISCUSSIONS AND CONCLUSIONS}

One way of attempting to help testing the standard accretion disk
models is to find relationships between variability and
fundamental parameters of AGNs, such as black hole mass. The
discovery of such a relationship -- or lack thereof -- may provide
useful clues to the physical mechanisms behind the variability.
Different sources of optical variations can be associated with
different characteristic time scales, and many of these time
scales depend on black hole mass. Wold et al. (2007) investigate
the dependence of quasar variability on black hole mass, and find
that the measured variability amplitude correlates with black hole
mass. Collier \& Peterson (2001) attempted to define a
relationship between black hole mass and characteristic
variability time scale. They reported evidence of black hole
masses correlating with characteristic optical variability time
scales for a sample of 10 well-monitored AGNs. In this paper, a
linear correlation between the measured time scales of optical
variations and the black hole masses is found for a sample of 26
well-monitored AGNs by reverberation mapping. This linear
correlation supports suggestion of Collier \& Peterson (2001). The
slope of this correlation in equation (15) is $\sim 0.12$, and
this slope is consistent with the one of $\sim 0.15$ predicted by
equation (3) with the viscosity parameter $\alpha=0.1$ and the
emitting radius $r_{\rm{d}}= 100 \/\ r_{\rm{g}}$. The slopes
between the thermal time scale and the black hole mass are
estimated for another two emitting radii of $r_{\rm{d}}= 50 \/\
r_{\rm{g}}$ and $r_{\rm{d}}= 200 \/\ r_{\rm{g}}$ in equation (3)
with $\alpha=0.1$. The three theoretical lines between time scales
and black hole masses are presented in Figure $6b$. It can be seen
in Figure $6b$ that the theoretical line of $r_{\rm{d}}= 100 \/\
r_{\rm{g}}$ matches the observed data points and the best fit line
better than the other two lines do. This means that the measured
characteristic time scales of optical variations are likely to be
from the accretion disk thermal instability. Then the standard
accretion disk models are likely to be conditionally favored by
observations.

Another way of attempting to help testing the standard accretion
disk models is connect the observed variability time scale with
one of the physical time scales in equations (1)--(4). Among the
four physical time scales, the thermal time scale is most close to
the measured optical variability time scale as $\alpha =0.1$. The
viscosity parameter $\alpha$ has the typical value of $\sim 0.1$
for the standard accretion disks (Shakura \& Sunyaev 1973). A
value of $\alpha \lesssim 1/2$ is implied by the condition that
the turbulence should be subsonic in the standard disks (Merloni
2003). A lower value of $\alpha < 0.14$ is suggested by numerical
investigations for thin accretion disks with a constant effective
speed of sound (Afshordi \& Paczynski 2003). Merloni \& Nayakshin
(2006) also limited a similar range of $\alpha \lesssim 0.15$ on
the basis of studying the limit-cycle instability in magnetized
accretion disks. Khajenabi \& Shadmehri (2007) adopted $\alpha
\sim 0.03$--$0.3$ to study the dynamical structure of a
self-gravitating disk. Then the viscosity paramter $\alpha$ in the
standard disks possibly has a wide range including the typical
value of $\alpha \sim 0.1$. If the viscosity parameter $\alpha$ is
allowed to range from $\sim 0.03$ to $\sim 0.2$, the time scales
of optical variations are consistent with the thermal time scales
predicted by the standard accretion disk models. This implies that
the measured characteristic time scales of optical variations are
likely to be produced by the accretion disk thermal instability.

The analysis shows that the wavelength differences $\Delta
\lambda$ are correlated with the relevant time lags between
different bands for NGC 7469, but there is no correlation between
the two quantities for NGC 5548. If the flux variations are caused
by the accretion flow fluctuations travelling inwards across the
emitting regions, it is likely that the shorter wavelength
variations lag the longer wavelength variations. However, the
longer wavelength variations lag the shorter wavelength variations
for NGC 4151 and NGC 7469. The shorter wavelength variations lag
the longer wavelength variations for NGC 5548 except that the
variations at 2787 $\rm{\AA}$ lag those at 1841 $\rm{\AA}$. The
variations at 1390 $\rm{\AA}$ lag those at 1880 $\rm{\AA}$ for
Fairall 9. If the optical/UV fluxes are the reprocessed continuum
with harder photons from the center of accretion disk and softer
ones at radii farther out, it is expected that the longer
wavelength variations lag the shorter wavelength variations.
However, the longer wavelength variations lead the shorter
wavelength variations for NGC 5548 and Fairall 9, and this is
inconsistent with the expectation in the case of continuum
reprocessing. The calculations for NGC 7469 and NGC 4151 show that
the time lags estimated in the case of continuum reprocessing are
marginally consistent with the measured time lags. In addition,
Fairall 9 and NGC 5548 have the black hole mass
$M_{\rm{BH}}>5\times 10^7\/\ \rm{M_{\sun}}$ with the longer
wavelength variations leading the shorter wavelength variations,
but NGC 4151 and NGC 7469 have $M_{\rm{BH}}<5\times 10^7\/\
\rm{M_{\sun}}$ with the shorter wavelength variations leading the
longer wavelength variations (see Table 3). There seems to be a
trend between black hole mass and time lag. As black hole mass is
above some value, the longer wavelength variations might lead the
shorter wavelength variations, but black hole mass is below this
value, the shorter wavelength variations might lead the longer
wavelength variations.

The origin of the radiation energies emitted in accretion disk is
a key to the issue that the harder photons lead or lag the softer
ones. For non-blazar type objects, if the optical/UV radiations
are the reprocessed X-rays that are commonly attributed to Compton
up-scattering of thermal UV seed photons by hot electrons in a
corona (e.g., Sunyaev \& Titarchuck 1980; Haardt \& Maraschi
1991), the opical/UV and X-ray light curves are expected to be
correlated with the X-rays leading the opitcal/UV radiation, and
then the harder and softer photons in optical--UV regime are
correlated with the harder photons leading the softer ones. The
opitcal/UV emissions in NGC 7469 and NGC 4151 probably belong to
this case. If the bulk of the observed optical/UV continuum arises
from the viscous dissipation in accretion disk, the resulting
light curves would be correlated but the UV radiations should lead
the X-rays. This scenario is supported by observations of the
Seyfert galaxy MCG--6-30-15 (Ar\'evalo et al. 2005). In this
scenario, the observed UV and the seed-photon-emitting regions are
connected by perturbations of the accretion flow travelling
inwards through the accretion disk, affecting first the main UV
emitting radii and then the innermost region where the bulk of the
seed photons is expected to be produced (e.g., Ar\'evalo et al.
2005). We analyzed the flux variations in 1--2 KeV (Leighly et al.
1997) and 1855 $\rm{\AA}$ (O'Brien et al. 1998) for 3C 390.3, and
a similar behavior to MCG--6-30-15 is found. The time lag
estimated by the ZDCF centroid for 3C 390.3 is
$\tau^{\rm{ob}}_{\rm{cent}}=-4.01^{-1.28}_{+0.77} \/\ \rm{days}$
with the X-rays lagging the UV radiation. The UV radiation emitted
by NGC 5548 and Fairall 9 might belong to the thermal radiation
from the viscous dissipation, and perturbations of the accretion
flow travelling inwards through the accretion disk result in the
softer photons leading the harder ones. Our results may support
that the signs of time lags differ from case to case (e.g., Maoz
et al. 2002). The existences of negative as well as positive time
lags imply that different processes could be dominating the
emissions at different cases, and generally don't indicate any
simple relation between the energy bands.

In this paper, a sample of 26 objects well observed for
reverberation mapping is used to test the standard accretion disk
models accepted widely by comparing the theoretical expectations
to the measured time scales of optical variations, the observed
relation of the black hole masses with the measured time scales,
and the measured time lags between the optical/UV bands. The time
scales measured by both the most common definition of the
variability time scale and the zero-crossing time of the ACF are
consistent with each other (see Fig. 1). The observed variability
time scales are linearly correlated with the black hole masses
(see Fig. 6), and this linear relation is conditionally consistent
with expectation for the thermal time scales and the black hole
masses in equation (3). Adopting the viscosity parameter typically
of $\alpha \sim 0.1$ (Shakura \& Sunyaev 1973), the thermal time
scales are most close to the measured time scales of optical
variations. The combinations of $\alpha \sim 0.03$--$0.2$ and
$a_{\ast}=0.5$--$0.998$ could result in the thermal time scales
that are in well agreement with the optical variability time
scales presented in Table 2. Then it is likely that the optical
variations result from the accretion disk thermal instability. The
time lags are measured by the ZDCF method for four ones out of
these 26 objects. The analyzed results show that the harder and
softer photons at the optical/UV bands are correlated with the
harder photons leading the softer ones for NGC 4151 and NGC 7469,
and with the harder photons lagging the softer ones for NGC 5548
and Fairall 9 (see Table 3). For NGC 7469 and NGC 4151, the
measured time lags are marginally consistent with the time lags
estimated in the case of continuum thermal reprocessing. It is
possible that the optical/UV emissions of NGC 4151 and NGC 7469
are the reprocessed X-rays that are commonly attributed to Compton
up-scattering of thermal UV seed photons by hot electrons. For NGC
5548 and Fairall 9, the UV photons are unlikely to be from the
continuum thermal reprocessing in the accretion disk. Our
investigations on the variability time scales, the relation of the
variability time scales with the black hole masses, and the time
lags between different bands are unlikely to be inconsistent with
or are likely to be conditionally in favor of the standard
accretion disk models of AGNs.

\acknowledgements
We are grateful to the anonymous referee for
his/her constructive comments and suggestions leading to
significant improvement of this paper. We are also grateful to
Prof. S. Kaspi for his constructive comments and suggestions that
helped to significantly improve this paper. Prof. T. Alexander is
thanked for kindly providing his ZDCF code. HTL thanks for
financial support by National Natural Science Foundation of China
(Grant 10778702). JMB is supported by NSFC (Grant 10573030) and
(Grant 10778726).

\clearpage

\begin{deluxetable}{lrrrrr}
\tablenum{1} \tablewidth{0pt} \tablecaption{Sample and data}

\tablehead{\colhead{Objects}&\colhead{$z$}&\colhead{$\lambda(\rm{\AA})$}&\colhead{Refs.}
&\colhead{$\frac{M_{\rm{BH}}}{10^8\/\
M_{\sun}}$}&\colhead{$\frac{L_{\rm{bol}}}{\rm{ergs
\/\ s^{-1}}}$}\\
\colhead{(1)}&\colhead{(2)}&\colhead{(3)}&\colhead{(4)}&\colhead{(5)}&\colhead{(6)}}

\startdata

PG 0026+129&0.142&5100&1, 2&3.93&45.39\\
PG 0052+251&0.155&5100&1, 2&3.69&45.93\\
Fairall 9&0.047&5340&3&2.55&45.23\\
         &     &1880&4& &  \\
         &     &1390&4& &  \\
PG 0804+761&0.100&5100&1, 2&6.93&45.93\\
PG 0844+349&0.064&5100&1, 2&9.24&45.36\\
PG 0953+414&0.239&5100&1, 2&2.76&46.16\\
NGC 3783&0.010&5150&5&0.30&44.41\\
NGC 4051&0.002&5100&6&0.02&43.56\\
NGC 4151&0.003&5125&7&0.13&43.73\\
        &     &2688&8& & \\
        &     &1440&8 & & \\
        &     &1275&8 & & \\
PG 1211+143&0.085&5100&1, 2&1.46&45.81\\
PG 1226+023&0.158&5100&1, 2&8.86&47.35\\
PG 1229+204&0.064&5100&1, 2&0.73&45.01\\
PG 1307+085&0.155&5100&1, 2&4.40&45.83\\
Mrk 279&0.030&5100&9&0.35&44.83\\
PG 1351+640&0.087&5100&1, 2&0.46&45.50\\
PG 1411+442&0.089&5100&1, 2&4.43&45.58\\
NGC 5548&0.017&5150&10&0.67&44.83\\
        &     &2787&11& & \\
        &     &2441&11& & \\
        &     &2237&11& & \\
        &     &1841&11& & \\
        &     &1749&11& & \\
        &     &1378&11& & \\
PG 1426+015&0.086&5100&1, 2&12.98&45.19\\
PG 1613+658&0.129&5100&1, 2&2.79&45.66\\
PG 1617+175&0.114&5100&1, 2&5.94&45.22\\
PG 1700+518&0.292&5100&1, 2&7.81&46.56\\
PG 1704+608&0.371&5100&1, 2&0.37&46.33\\
3C 390.3&0.056&5177&12&2.87&44.88\\
Mrk 509&0.034&5100&13&1.43&45.03\\
PG 2130+099&0.061&5100&1, 2&4.57&45.47\\
NGC 7469&0.016&4845&14&0.12&45.28\\
        &     &6962&14& & \\
        &     &1825&15& & \\
        &     &1740&15& & \\
        &     &1485&15& & \\
        &     &1315&15& & \\
\enddata
\tablecomments{Col: (1) name. Col: (2) redshift. Col: (3) the rest
frame wavelengths of light curves. Col: (4) the references of
light curves. Col: (5) black hole mass. Col: (6) log of the
bolometric luminosity.}
\tablerefs{(1) Kaspi et al. 2000; (2)
Kaspi et al. 2005; (3) Santos-Lleo et al. 1997; (4)
Rodriguez-Pascual et al. 1997;(5) Stirpe et al. 1994; (6) Peterson
et al. 2000; (7) Kaspi et al. 1996; (8) Crenshaw et al. 1996; (9)
Santos-Lleo et al. 2001; (10) Wanders \& Peterson 1996; (11) the
UltraViolet Light Curve Database for AGNs; (12) Dietrich et al.
1998; (13) Carone et al. 1996; (14) Collier et al. 1998; (15)
Kriss et al. 2000.}
\end{deluxetable}

\clearpage

\begin{deluxetable}{lrrrrrr}
\tablenum{2} \tablewidth{0pt} \tablecaption{Calculated results}

\tablehead{\colhead{Objects}&\colhead{$\frac{\tau
/(1+z)}{\rm{days}}$}&\colhead{$\frac{\tau_{\rm{0}}
/(1+z)}{\rm{days}}$}&\colhead{$\frac{\tau_{\rm{l}}}{\rm{days}}$}&\colhead{$\frac{\tau_{\rm{dyn}}}{\rm{days}}$}
&\colhead{$\frac{\tau_{\rm{th}}}{\rm{days}}$}&\colhead{$\frac{\tau_{\rm{s}}}{\rm{days}}$}\\
\colhead{(1)}&\colhead{(2)}&\colhead{(3)}&\colhead{(4)}&\colhead{(5)}&\colhead{(6)}&\colhead{(7)}}

\startdata

PG 0026&$343.6\pm 1.6$&$182.7^{-3.5}_{+9.2}$&1.20&8.22&84.01&2.14 $10^4$\\
       &     &     &1.02&6.39&65.74&1.82 $10^4$\\
       &     &     &0.99&6.08&62.09&1.78 $10^4$\\
       &     &     &0.79&4.26&43.83&1.41 $10^4$\\
PG 0052&$242.5\pm 1.8$&$326.4^{-6.4}_{+10.5}$&1.85&16.44&164.36&3.31 $10^4$\\
       &     &     &1.64&13.70&135.14&2.94 $10^4$\\
       &     &     &1.52&12.18&120.53&2.72 $10^4$\\
       &     &     &1.20&8.52&84.01&2.15 $10^4$\\
Fairall 9&$160.8\pm 0.3$&$110.6^{-2.6}_{+8.8}$&1.00&7.91&76.70&1.83 $10^4$\\
         &     &     &0.89&6.39&65.75&1.62 $10^4$\\
         &     &     &0.82&5.78&58.44&1.51 $10^4$\\
         &     &     &0.65&3.96&40.18&1.19 $10^4$\\
PG 0804&$385.5\pm 0.7$&$497.0^{-5.7}_{+7.2}$&2.20&15.52&153.41&3.93 $10^4$\\
       &     &     &1.97&13.09&131.49&3.52 $10^4$\\
       &     &     &1.83&11.57&116.88&3.27 $10^4$\\
       &     &     &1.45&8.22&84.01&2.59 $10^4$\\
PG 0844&$313.1\pm 1.3$&$134.8^{-5.3}_{+9.0}$&1.44&7.00&69.40&2.57 $10^4$\\
       &     &     &1.31&6.09&62.09&2.34 $10^4$\\
       &     &     &1.23&5.48&54.79&2.19 $10^4$\\
       &     &     &0.99&3.96&40.18&1.76 $10^4$\\
PG 0953&$275.1\pm 0.7$&$772.0^{-5.2}_{+6.3}$&2.05&21.92&219.15&3.67 $10^4$\\
       &     &     &1.82&18.26&182.63&3.25 $10^4$\\
       &     &     &1.68&16.44&164.36&3.01 $10^4$\\
       &     &     &1.35&11.57&113.23&2.37 $10^4$\\
NGC 3783&$38.2\pm 2.8$&$29.5^{-0.14}_{+0.20}$&0.26&3.04&29.22&4.69 $10^3$\\
        &    &     &0.23&2.44&25.57&4.15 $10^3$\\
        &    &     &0.21&2.13&21.92&3.84 $10^3$\\
        &    &     &0.17&1.52&14.61&3.02 $10^3$\\
NGC 4051&$120.3\pm 0.05$&$165.4^{-0.40}_{+0.30}$&0.06&1.22&10.96&1.02 $10^3$\\
        &     &     &0.05&0.91&10.96&8.99 $10^2$\\
        &     &     &0.05&0.91&7.31&8.29 $10^2$\\
        &     &     &0.04&0.61&7.31&6.50 $10^2$\\
NGC 4151&$70.4\pm 0.6$&$45.0^{-0.20}_{+0.13}$&0.12&1.52&14.61&2.10 $10^3$\\
        &    &     &0.10&1.22&10.96&1.85 $10^3$\\
        &    &     &0.09&0.91&10.96&1.71 $10^3$\\
        &    &     &0.08&0.61&7.31&1.35 $10^3$\\
PG 1211&$237.1\pm 0.8$&$533.4^{-3.8}_{+11.9}$&1.28&14.91&149.75&2.29 $10^4$\\
       &     &     &1.13&12.48&124.19&2.03 $10^4$\\
       &     &     &1.05&10.96&109.58&1.87 $10^4$\\
       &     &     &0.83&7.61&76.70&1.47 $10^4$\\
PG 1226&$314.4\pm 5.7$&$401.2^{-2.6}_{+5.5}$&7.61&87.66&876.60&1.36 $10^5$\\
       &     &     &6.72&72.75&726.85&1.20 $10^5$\\
       &     &     &6.23&64.83&650.15&1.11 $10^5$\\
       &     &     &4.90&45.35&452.91&8.77 $10^4$\\
PG 1229&$124.5\pm 4.3$&$143.0^{-2.2}_{+2.8}$&0.55&5.78&58.44&9.76 $10^3$\\
       &     &     &0.48&4.87&47.48&8.62 $10^3$\\
       &     &     &0.44&4.26&40.18&7.99 $10^3$\\
       &     &     &0.35&3.04&29.22&6.30 $10^3$\\
PG 1307&$234.2\pm 7.7$&$294.8^{-3.6}_{+4.3}$&1.79&14.31&142.45&3.20 $10^4$\\
       &     &     &1.59&11.87&120.53&2.84 $10^4$\\
       &     &     &1.48&10.65&105.92&2.64 $10^4$\\
       &     &     &1.17&7.61&76.70&2.09 $10^4$\\
Mrk 279&$86.5\pm 2.9$&$76.5^{-0.15}_{+0.91}$&0.38&4.87&47.48&6.78 $10^3$\\
       &    &     &0.33&3.96&40.18&5.98 $10^3$\\
       &    &     &0.31&3.65&36.53&5.53 $10^3$\\
       &    &     &0.24&2.44&25.57&4.35 $10^3$\\
PG 1351&$423.6\pm 1.7$&$476.4^{-2.8}_{+4.9}$&0.70&10.96&109.58&1.26 $10^4$\\
       &     &     &0.62&9.13&91.31&1.11 $10^4$\\
       &     &     &0.57&7.91&76.70&1.02 $10^4$\\
       &     &     &0.45&5.48&54.79&8.07 $10^3$\\
PG 1411&$317.4\pm 1.4$&$356.6^{-2.3}_{+4.1}$&1.45&10.35&102.27&2.60 $10^4$\\
       &     &     &1.30&8.83&87.66&2.32 $10^4$\\
       &     &     &1.20&7.61&76.70&2.16 $10^4$\\
       &     &     &0.96&5.48&54.79&1.71 $10^4$\\
NGC 5548&$45.1\pm 1.1$&$141.5^{-2.8}_{+7.1}$&0.47&4.87&47.48&8.36 $10^3$\\
        &    &     &0.41&3.96&40.18&7.37 $10^3$\\
        &    &     &0.38&3.35&36.53&6.86 $10^3$\\
        &    &     &0.30&2.44&25.57&5.38 $10^3$\\
PG 1426&$638.7\pm 1.9$&$852.0^{-18.0}_{+20.2}$&1.30&5.17&51.14&2.33 $10^4$\\
       &     &     &1.22&4.57&47.48&2.17 $10^4$\\
       &     &     &1.15&4.26&43.83&2.06 $10^4$\\
       &     &     &0.94&3.04&32.87&1.68 $10^4$\\
PG 1613&$309.5\pm 1.2$&$232.1^{-4.9}_{+7.5}$&1.37&11.87&120.53&2.45 $10^4$\\
       &     &     &1.22&10.04&98.62&2.17 $10^4$\\
       &     &     &1.12&8.83&87.66&2.01 $10^4$\\
       &     &     &0.89&6.39&62.09&1.59 $10^4$\\
PG 1617&$277.9\pm 3.6$&$187.9^{-6.3}_{+6.1}$&1.15&6.39&62.09&2.06 $10^4$\\
       &     &     &1.04&5.48&54.79&1.86 $10^4$\\
       &     &     &0.97&4.87&47.48&1.73 $10^4$\\
       &     &     &0.77&3.35&36.53&1.38 $10^4$\\
PG 1700&$470.4\pm 87.4$&$419.3^{-10.2}_{+12.2}$&3.85&33.48&336.03&6.88 $10^4$\\
       &     &     &3.42&28.00&281.24&6.11 $10^4$\\
       &     &     &3.17&24.96&252.02&5.68 $10^4$\\
       &     &     &2.50&17.65&175.32&4.47 $10^4$\\
PG 1704&$286.7\pm 5.9$&$389.1^{-7.4}_{+8.8}$&1.27&29.22&292.20&2.27 $10^4$\\
       &     &     &1.12&24.05&241.07&1.99 $10^4$\\
       &     &     &1.03&21.31&211.85&1.84 $10^4$\\
       &     &     &0.81&14.91&149.75&1.45 $10^4$\\
3C 390.3&$143.8\pm 0.6$&$50.6^{-0.46}_{+0.63}$&0.73&4.57&47.48&1.32 $10^4$\\
        &     &     &0.66&3.96&40.18&1.23 $10^4$\\
        &     &     &0.61&3.35&32.87&1.10 $10^4$\\
        &     &     &0.49&2.44&25.57&8.76 $10^3$\\
Mrk 509&$258.7\pm 1.0$&$181.0^{-1.7}_{+2.2}$&0.67&5.78&58.44&1.20 $10^4$\\
       &     &     &0.60&4.87&47.48&1.07 $10^4$\\
       &     &     &0.55&4.26&40.18&9.92 $10^3$\\
       &     &     &0.44&3.04&29.22&7.84 $10^3$\\
PG 2130&$319.0\pm 12.3$&$332.9^{-5.0}_{+11.0}$&1.28&8.52&84.01&2.28 $10^4$\\
       &     &     &1.20&7.61&76.70&2.14 $10^4$\\
       &     &     &1.11&6.70&69.40&1.99 $10^4$\\
       &     &     &0.88&4.87&47.48&1.58 $10^4$\\
NGC 7469&$74.9\pm 0.8$&$4.5^{-0.05}_{+0.02}$&0.36&7.91&80.36&6.33 $10^3$\\
        &    &     &0.32&6.39&65.75&5.56 $10^3$\\
        &    &     &0.29&5.48&54.79&5.13 $10^3$\\
        &    &     &0.23&4.26&40.18&4.16 $10^3$\\
\enddata
\tablecomments{Col: (1) name. Col: (2) variability time scale at
the optical band. Col: (3) variability time scale obtained by the
zero-crossing time of the ACF estimated by the ZDCF. Col: (4)
light crossing time scales. Col: (5) dynamical time scales. Col:
(6) thermal time scales. Col: (7) sound crossing time scales. For
each object, the first, second, third, and forth values listed in
columns (4)--(7) are calculated from the standard accretion disks
under the Kerr metric with spin parameter $a_{\ast}=0.5$, $0.8$,
$0.9$, and $0.998$, respectively.}
\end{deluxetable}

\clearpage

\begin{deluxetable}{lrrrrr}
\tablenum{3} \tablewidth{0pt} \tablecaption{Time lags}

\tablehead{

\colhead{Objects}&\colhead{$\lambda_1(\rm{\AA})$}&\colhead{$\lambda_2(\rm{\AA})$}

&\colhead{$\tau^{\rm{ob}}_{\rm{cent}}(\rm{days})$}&\colhead{$\tau_{\rm{rep}}(\rm{days})$}
&\colhead{$\tau_{\rm{lag}}(\rm{yrs})$}\\

\colhead{(1)}&\colhead{(2)}&\colhead{(3)}&\colhead{(4)\tablenotemark{a}}&
\colhead{(5)\tablenotemark{a}}&\colhead{(6)\tablenotemark{a}}

}

\startdata

Fairall 9&1390&1880&$-3.71^{-0.66}_{+0.68}$&$\tablenotemark{b} \/\ 0.08 \/\ 0.07 \/\ 0.05$&-1.95 \\

NGC 4151&1440&2688&$0.17^{-0.01}_{+0.08}$&$0.03 \/\ 0.02 \/\ 0.02 \/\ 0.02$&-0.75\\
        &1275& 2688&$0.09^{-0.01}_{+0.08}$&$0.03 \/\ 0.03 \/\ 0.03 \/\ 0.02$&-0.82\\

NGC 5548&1841&2787&$0.72^{-0.16}_{+0.18}$&$0.09 \/\ 0.08 \/\ 0.07 \/\ 0.06$& -2.52\\
        &1841&2441&$-0.54^{-0.15}_{+0.17}$&$0.06 \/\ 0.05 \/\ 0.04 \/\ 0.03$& -1.49\\
        &1841&2237&$-0.55^{-0.16}_{+0.17}$&$0.04 \/\ 0.03 \/\ 0.03 \/\ 0.02$&-0.93 \\
        &1749&2787&$-0.98^{-0.11}_{+0.17}$&$0.10 \/\ 0.09 \/\ 0.08 \/\ 0.06$& -2.72\\
        &1749&2441&$-0.67^{-0.20}_{+0.21}$&$0.06 \/\ 0.06 \/\ 0.05 \/\ 0.04$& -1.69\\
        &1749&2237&$-0.68^{-0.19}_{+0.20}$&$0.04 \/\ 0.04 \/\ 0.03 \/\ 0.03$& -1.13\\
        &1378&2787&$-0.64^{-0.18}_{+0.19}$&$0.13 \/\ 0.11 \/\ 0.10 \/\ 0.08$&-3.41 \\
        &1378&2441&$-0.61^{-0.18}_{+0.19}$&$0.09 \/\ 0.08 \/\ 0.07 \/\ 0.06$& -2.39 \\
        &1378&2237&$-0.64^{-0.18}_{+0.19}$&$0.07 \/\ 0.06 \/\ 0.06 \/\ 0.05$&-1.82 \\
        &1378&1841&$-1.43^{-0.05}_{+0.06}$&$0.04 \/\ 0.03 \/\ 0.03 \/\ 0.02$&-0.90 \\

NGC 7469&1825&6962&$0.83^{-0.04}_{+0.13}$&$0.50 \/\ 0.44 \/\ 0.41 \/\ 0.32$ &-19.88 \\
        &1740&6962&$1.46^{-0.08}_{+0.18}$&$0.51 \/\ 0.44 \/\ 0.41 \/\ 0.32$ &-20.03 \\
        &1485&6962&$1.72^{-0.08}_{+0.18}$&$0.52 \/\ 0.46 \/\ 0.42 \/\ 0.33$ &-20.45 \\
        &1315&6962&$1.72^{-0.08}_{+0.18}$&$0.53 \/\ 0.47 \/\ 0.43 \/\ 0.34$ & -20.70\\
        &1825&4845&$0.99^{-0.08}_{+0.17}$&$0.27 \/\ 0.24 \/\ 0.22 \/\ 0.17$ &-9.29 \\
        &1740&4845&$0.99^{-0.08}_{+0.17}$&$0.28 \/\ 0.24 \/\ 0.22 \/\ 0.18$ &-9.44 \\
        &1485&4845&$1.01^{-0.07}_{+0.16}$&$0.29 \/\ 0.26 \/\ 0.24 \/\ 0.19$ &-9.86\\
        &1315&4845&$1.23^{-0.08}_{+0.17}$&$0.30 \/\ 0.27 \/\ 0.25 \/\ 0.19$ &-10.10 \\
        &1315&1825&$0.32^{-0.19}_{+0.27}$&$0.03 \/\ 0.03 \/\ 0.03 \/\ 0.02$ &-0.82\\
\enddata

\tablenotetext{a}{The sign of the time lag is defined as
$\tau_{\rm{cent,rep,lag}}=t(\lambda_2)-t(\lambda_1$).}
\tablenotetext{b}{In case of $a_{\ast}=0.5$, the temperature of
disk is not high enough to emit radiation at $1390 \/\ \rm{\AA}$.}
\tablecomments{Col: (1) name. Col: (2) wavelengths of the first
light curves. Col: (3) wavelengths of the second light curves.
Col: (4) the observed time lags of the first light curves relative
to the second ones. Col: (5) the continuum reprocessing time lags,
and the first, second, third, and forth values are estimated from
the standard accretion disks under the Kerr metric with spin
parameter $a_{\ast}=0.5$, $0.8$, $0.9$, and $0.998$, respectively.
Col: (6) the sound crossing time lags estimated with spin
$a_{\ast}=0.998$.}
\end{deluxetable}

\clearpage

\begin{figure}
\centering
\includegraphics[angle=-90,scale=.5]{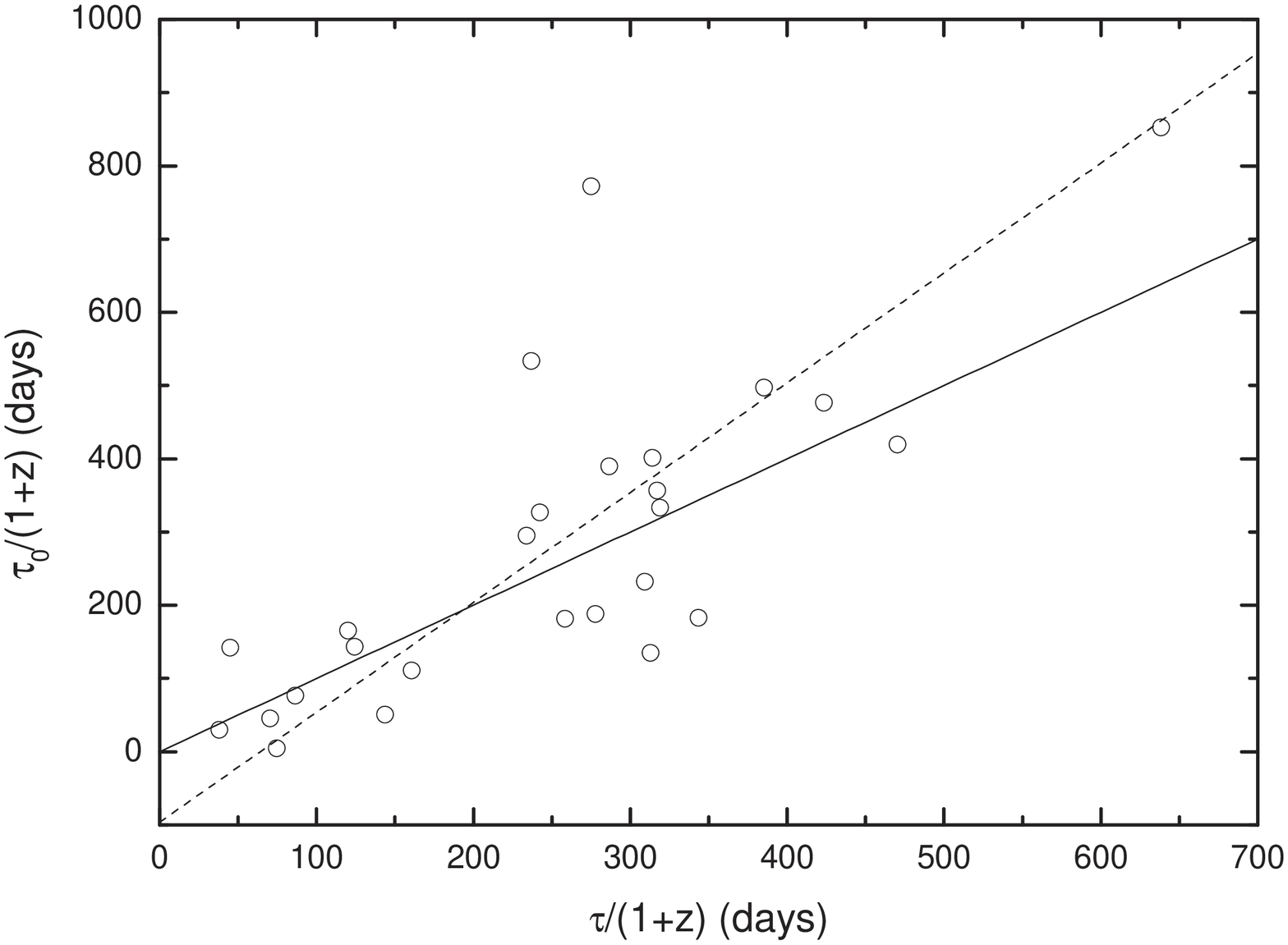}
 \caption{The plot of $\tau_0$ vs $\tau$. The solid line presents $\tau_0 =\tau$.
 The dashed line denotes the best linear fit.}
 \label{fig1}
\end{figure}

\clearpage

\begin{figure}
\centering
\includegraphics[angle=-90,scale=.5]{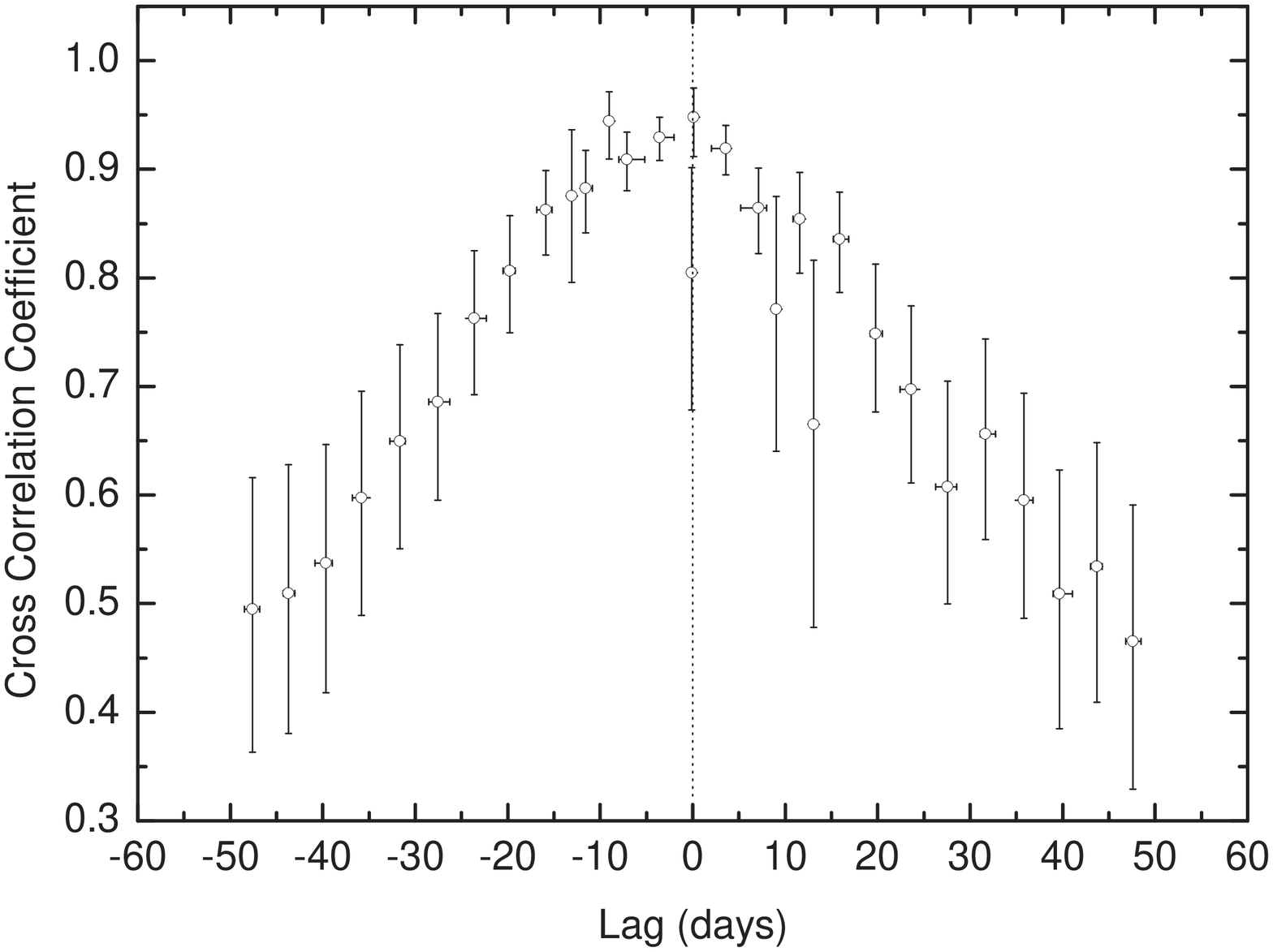}
 \caption{ZDCF for Fairall 9 between $1880\/\ \rm{\AA}$ and $1390\/\ \rm{\AA}$ light curves.}
 \label{fig2}
\end{figure}

\clearpage

\begin{figure}
\centering
\includegraphics[angle=-90,scale=.5]{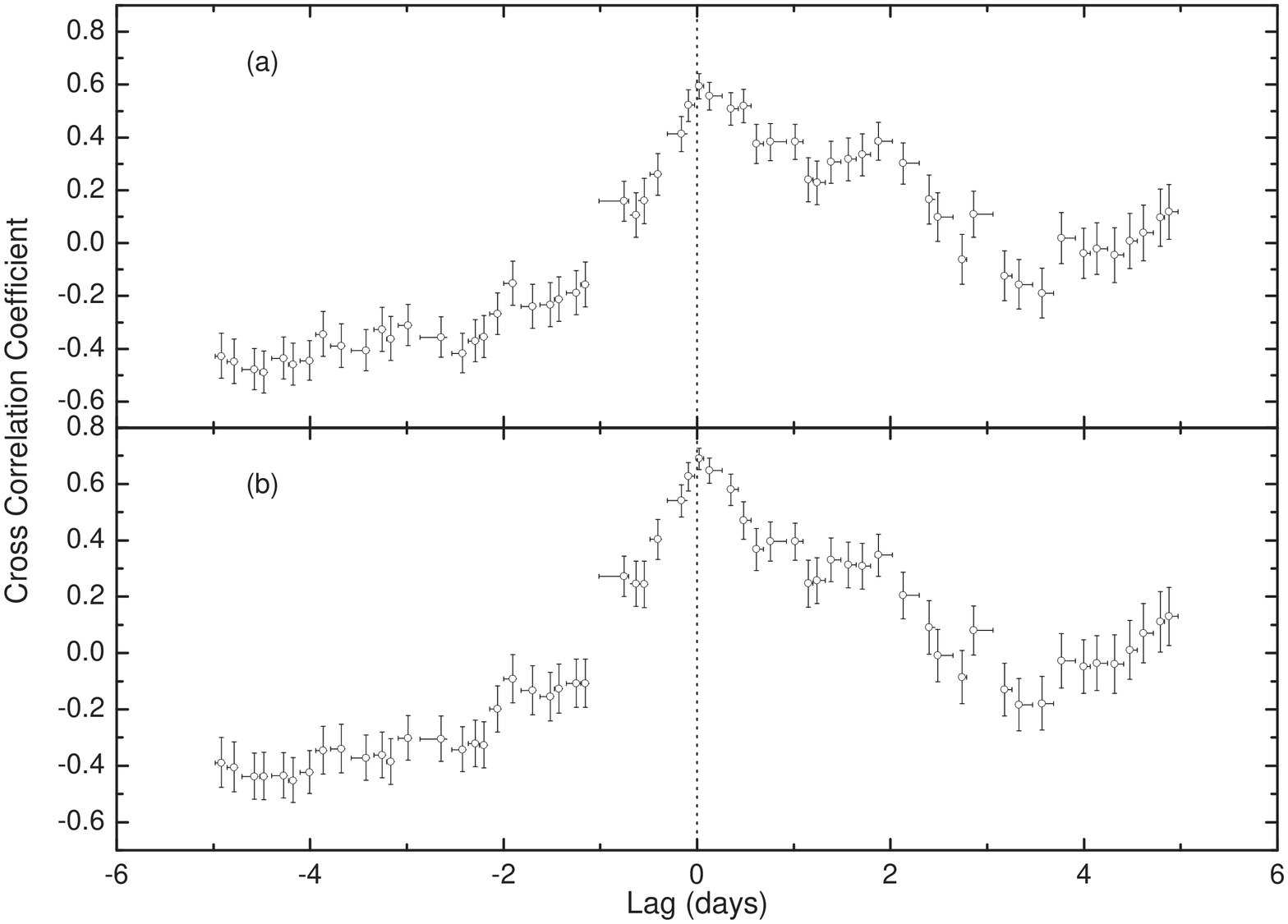}
 \caption{ZDCF for NGC 4151. ($a$) ZDCF between $2688\/\ \rm{\AA}$ and $1440\/\ \rm{\AA}$ light curves.
 ($b$) ZDCF between $2688\/\ \rm{\AA}$ and $1275\/\ \rm{\AA}$.}
 \label{fig3}
\end{figure}

\clearpage

\begin{figure}
\centering
\includegraphics[angle=-90,scale=.5]{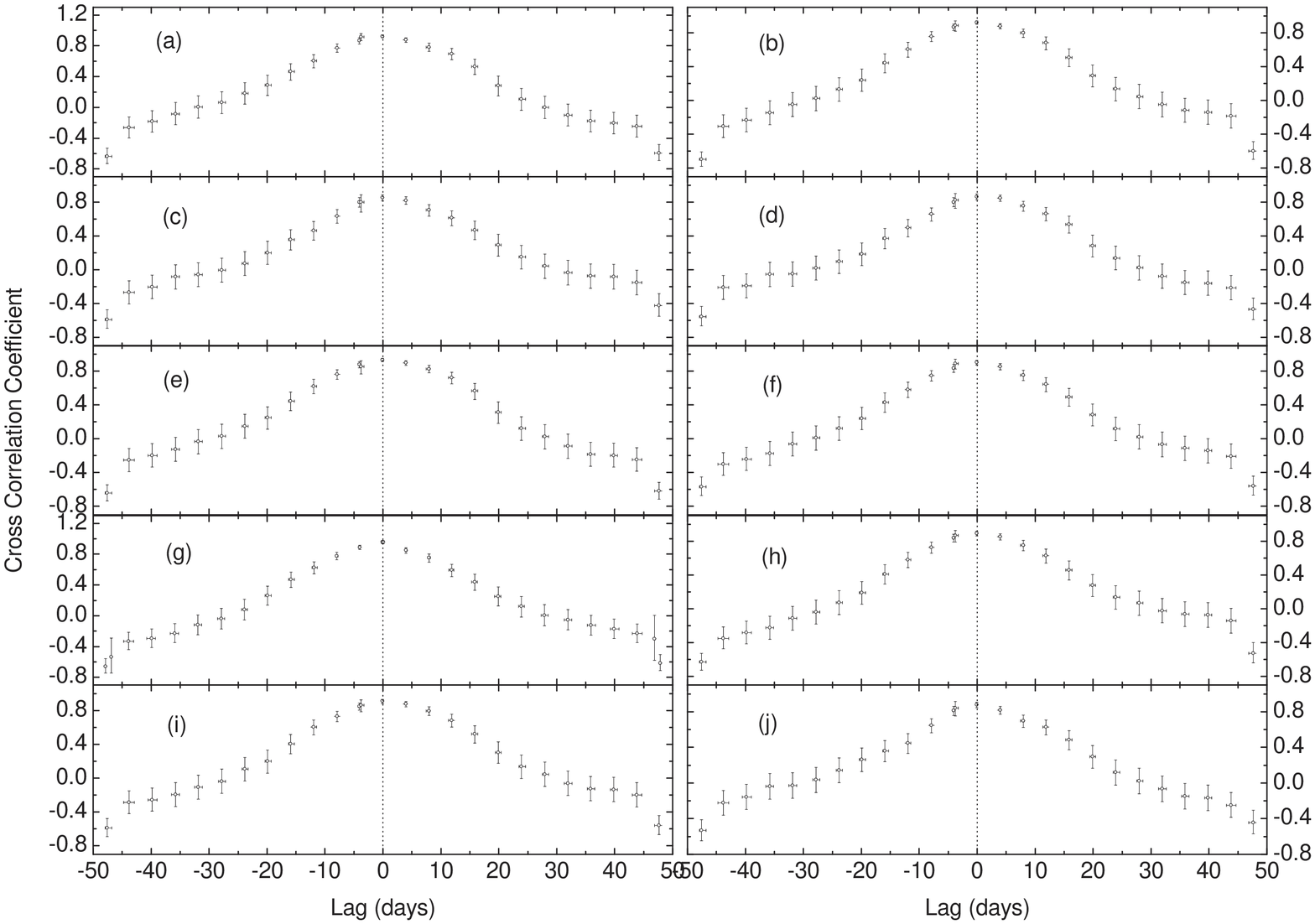}
 \caption{ZDCF for NGC 5548. ($a$) ZDCF between $2441\/\ \rm{\AA}$ and $1749\/\ \rm{\AA}$ light curves.
 ($b$) ZDCF between $2441\/\ \rm{\AA}$ and $1378\/\ \rm{\AA}$. ($c$) ZDCF between $2787\/\ \rm{\AA}$ and
 $1378\/\ \rm{\AA}$. ($d$) ZDCF between $2787\/\ \rm{\AA}$ and $1841\/\ \rm{\AA}$. ($e$) ZDCF between
 $2441\/\ \rm{\AA}$ and $1841\/\ \rm{\AA}$. ($f$) ZDCF between $2237\/\ \rm{\AA}$ and $1749\/\ \rm{\AA}$.
 ($g$) ZDCF between $1841\/\ \rm{\AA}$ and $1378\/\ \rm{\AA}$. ($h$) ZDCF between $2237\/\ \rm{\AA}$ and
 $1378\/\ \rm{\AA}$. ($i$) ZDCF between $2237\/\ \rm{\AA}$ and $1841\/\ \rm{\AA}$. ($j$) ZDCF between
 $2787\/\ \rm{\AA}$ and $1749\/\ \rm{\AA}$.}
 \label{fig4}
\end{figure}

\clearpage

\begin{figure}
\centering
\includegraphics[width=1\textwidth]{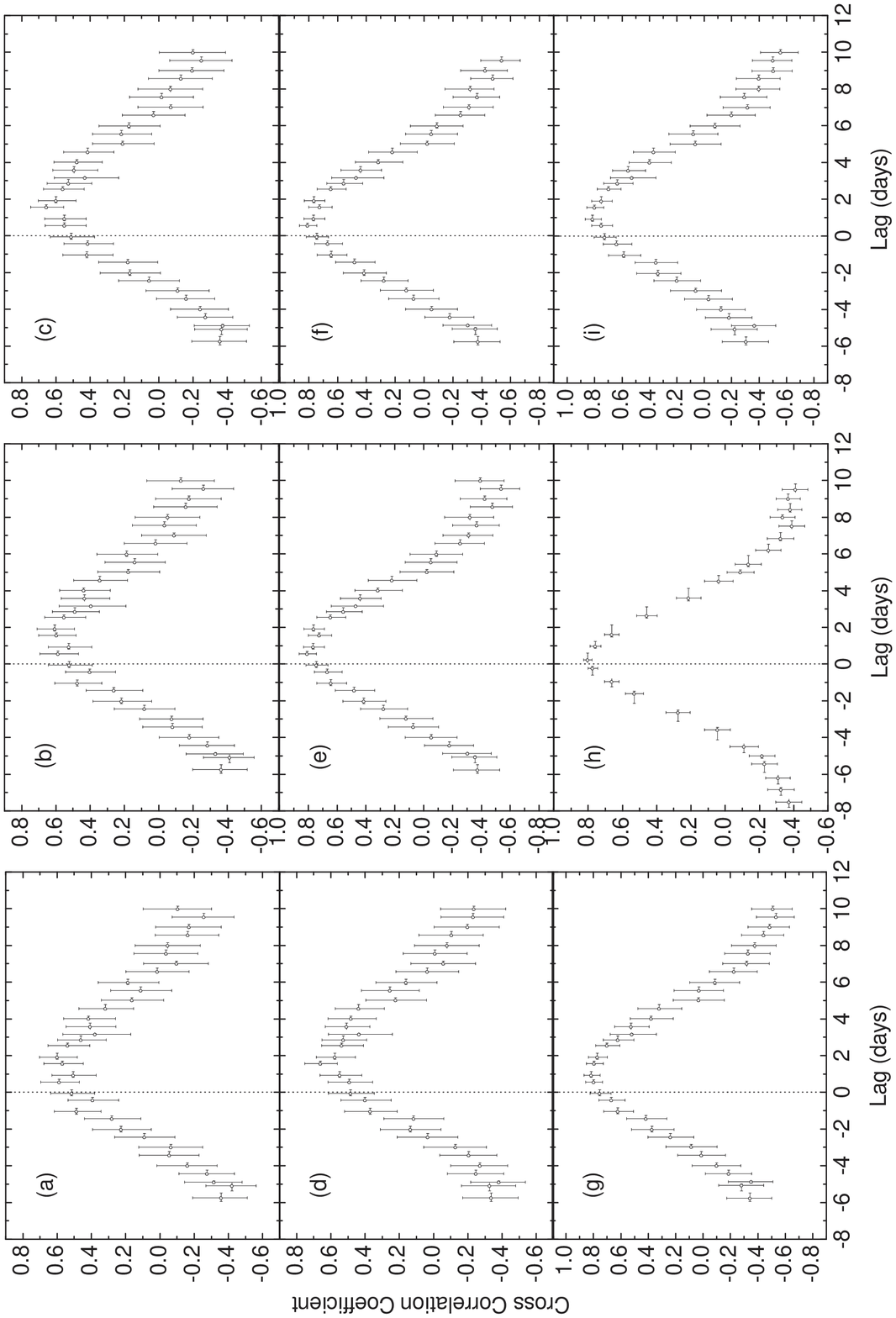}
 \caption{ZDCF for NGC 7469. ($a$) ZDCF between $6962\/\ \rm{\AA}$ and $1825\/\ \rm{\AA}$.
 ($b$) ZDCF between $6962\/\ \rm{\AA}$ and $1740\/\ \rm{\AA}$. ($c$) ZDCF between $6962\/\ \rm{\AA}$
 and $1485\/\ \rm{\AA}$. ($d$) ZDCF between $6962\/\ \rm{\AA}$ and $1315\/\ \rm{\AA}$.
 ($e$) ZDCF between $4845\/\ \rm{\AA}$ and $1825\/\ \rm{\AA}$. ($f$) ZDCF between $4845\/\ \rm{\AA}$
 and $1740\/\ \rm{\AA}$. ($g$) ZDCF between $4845\/\ \rm{\AA}$ and $1485\/\ \rm{\AA}$.
 ($h$) ZDCF between $1825\/\ \rm{\AA}$ and $1315\/\ \rm{\AA}$. ($i$) ZDCF between $4845\/\ \rm{\AA}$
 and $1315\/\ \rm{\AA}$.}
 \label{fig5}
\end{figure}

\clearpage

\begin{figure}
\centering
\includegraphics[angle=-90,scale=.5]{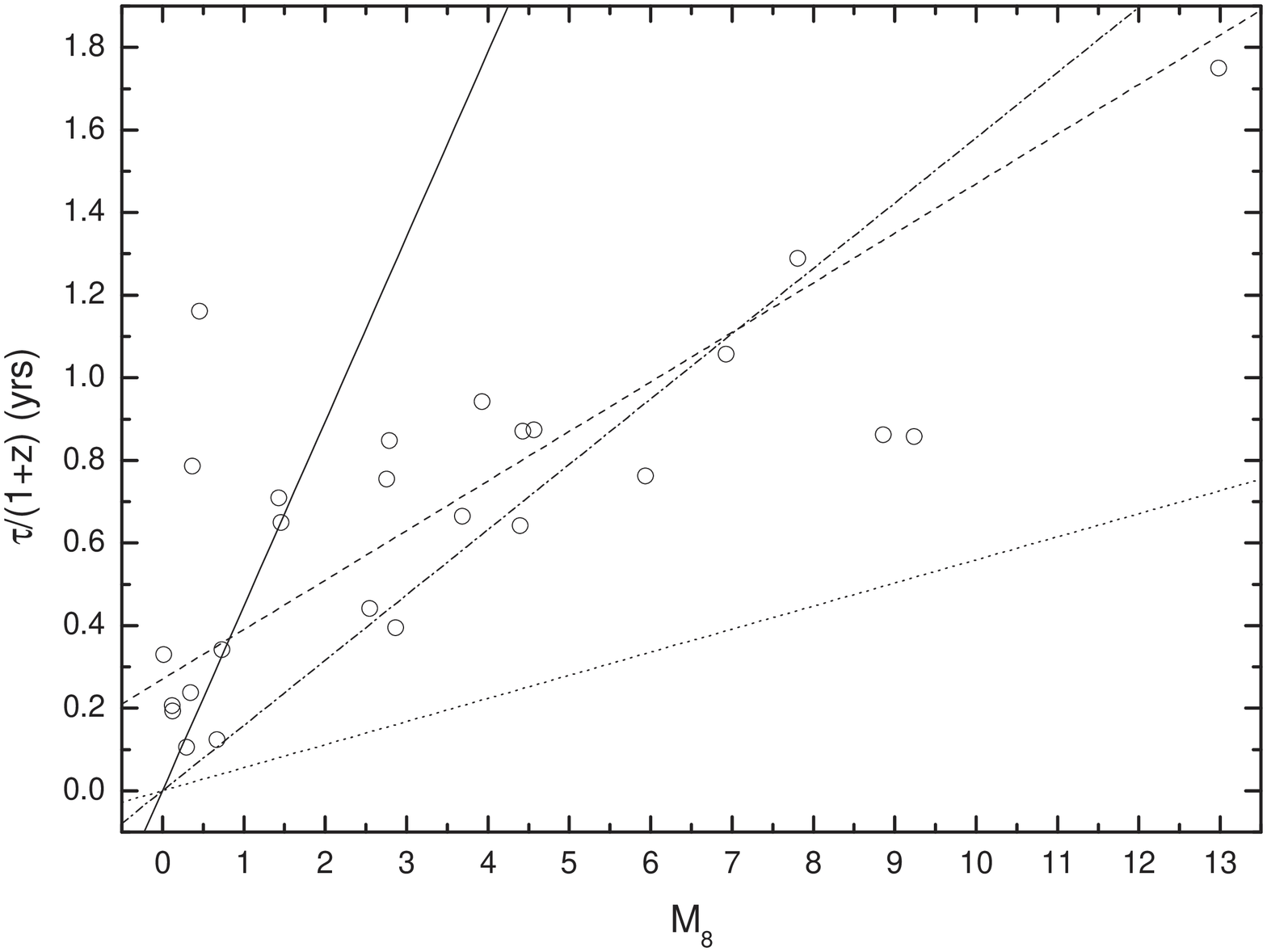}
 \caption{The plot of $\tau$ vs $M_{\rm{8}}$. The dashed line denotes the best linear fit.
 The dotted, dash-dotted, and solid lines are the theoretical lines of equation (3) for
 $r_{\rm{d}}=50 \/\ r_{\rm{g}}$, $100 \/\ r_{\rm{g}}$, and $200 \/\ r_{\rm{g}}$, respectively.}
\label{fig6}
\end{figure}

\end{document}